\documentclass[a4,11pt]{article}
\title{Assessing the accuracy of 
quantum Monte Carlo and density functional theory for energetics of small water clusters}
\author{M. J. Gillan$^{1,2,3}$, F. R. Manby$^{5}$, M. D. Towler$^{4,6,7}$ and 
D. Alf\`{e}$^{1,2,3,4}$ \\
$^1$London Centre for Nanotechnology, Gordon St, London WC1H 0AH, UK \\
$^2$Dept of Physics and Astronomy, UCL, Gower St, London WC1E 6BT, UK \\
$^3$Thomas Young Centre, UCL, Gordon St, London WC1H 0AH, UK \\
$^4$Dept of Earth Sciences, UCL, Gower St, London WC1E 6BT, UK \\
$^5$Centre for Computational Chemistry, School of Chemistry, University
of Bristol, \\
Bristol BS8 1TS, UK \\
$^6$TCM Group, Cavendish Laboratory, University of Cambridge, \\ 
Madingley Rd, Cambridge CB3 0HE, UK \\
$^7$Apuan Alps Centre for Physics, via del Collegio 22, Vallico Sotto \\ 
Fabbriche di Vallico 55020, Italy
}
\usepackage{graphicx,amsmath}

\newcommand\deltaone{\ensuremath{\Delta_1}}
\newcommand\deltatwo{\ensuremath{\Delta_{12}}}
\newcommand\deltathree{\ensuremath{\Delta_{123}}}

\begin{document}

\maketitle
\abstract{
We present a detailed study of the energetics of water 
clusters (H$_2$O)$_n$ with $n \le 6$, comparing diffusion Monte
Carlo (DMC) and approximate density functional theory (DFT) with well converged coupled-cluster
benchmarks.
We use the many-body decomposition
of the total energy to classify the errors of DMC and DFT
into 1-body, 2-body and beyond-2-body components.
Using both equilibrium cluster configurations and thermal
ensembles of configurations, we find DMC to be uniformly much more accurate than DFT, partly
because some of the approximate functionals give poor 1-body distortion energies.
Even when these are corrected, DFT remains considerably less accurate than DMC. 
When both 1- and 2-body errors of DFT
are corrected, some functionals compete in accuracy with DMC;
however, other functionals remain worse, showing that they
suffer from significant beyond-2-body errors. Combining the evidence
presented here with the recently demonstrated high accuracy of DMC
for ice structures, we suggest how DMC can now be used to provide
benchmarks for larger clusters and for bulk liquid water.
}

\section{Introduction}
\label{sec:intro}

Water in its many forms is one of the most intensively studied of all
substances, because of its relevance to so many different scientific
and technological fields. But in spite of many decades of effort, a fully
comprehensive account of the energetics of water systems at the molecular
level is still lacking. Density functional theory (DFT) is very
important in water studies, since it is readily applied to the bulk
liquid~\cite{laasonen93,tuckerman95,sprik96,silvistrelli99}
and solid~\cite{lee92,hamann97,singer05,dekoning06}
and their surfaces~\cite{kuo04,pan08}, 
as well as to solutions~\cite{marx97,tateyama05},
and to interfaces with other materials~\cite{liu09,liu10}. 
Unfortunately, DFT does not yet give satisfactory overall
accuracy~\cite{grossman04,allesch04,fernandez-serra04,vandevondele05,lee06,todorova06,guidon08}, 
and the past few years have seen strenuous
efforts to analyse and remedy the deficiencies of current
DFT approximations~\cite{jwang11,santra11,mogelhoj11}. 
Wavefunction-based molecular quantum chemistry techniques, particularly MP2
(second-order M\o{}ller-Plesset) and CCSD(T) (coupled-cluster singles and doubles with
perturbative triples), the latter often referred to as the ``gold standard'', can
reliably deliver much higher accuracy than DFT. These quantum chemistry
techniques have been extensively used to study water
clusters~\cite{xantheas94,xantheas95,pedulla96,burnham99,burnham02,klopper00,tschumper02,xu04,zhao05,anderson06,olson07,dahlke08,bates09,yoo10}
and to develop parameterised interaction
models~\cite{popkie73,matsuoka76,kim94a,hodges97,mas03,huang06,bukowski07,huang08,fanourgakis08,szalewicz09,kumar10,ywang11}.
However, the cost of obtaining this high accuracy increases dramatically
with system size. The direct application of correlated quantum chemistry methods to bulk 
solid and liquid water is still in its 
infancy~\cite{erba09,oneill11}. Recently, quantum Monte
Carlo (QMC) methods, and in particular diffusion Monte
Carlo (DMC)~\cite{hammond94,anderson99,nightingale99,foulkes01,needs10} 
have begun to be applied to both cluster and bulk
forms of water~\cite{santra11,gurtubay07,santra08} 
and the results indicate that DMC is much more
accurate than DFT for these systems as well as for weak non-covalent
interactions in other molecular systems~\cite{korth08,ma09}. Our purpose
in this paper is to present a systematic comparison of the accuracy of
DMC and DFT approximations for a wide range of configurations of 
small H$_2$O clusters, ranging from the monomer to the hexamer.
The use of thermal ensembles of configurations and the many-body
analysis of errors in the total energy are important themes of the work.

The errors of DFT are already troublesome for the H$_2$O monomer, and it
has been shown recently~\cite{santra09} 
that some common exchange-correlation functionals
give rather poor accuracy for its distortion energy, underestimating
the O-H bond-stretch energy. Difficulties with
the energetics of the dimer are well 
known~\cite{xantheas95,tsuzuki01}, 
with concerns about short-range
exchange-repulsion and intermediate-range hydrogen bonding, as well
as the lack of long-range dispersion in standard forms of DFT. Many-body
interactions are also known to play an important role in water 
energetics~\cite{xantheas94,pedulla96,bukowski07,fanourgakis08,szalewicz09,kumar10,hankins70,white90}, with 
induction effects due to the dipolar and higher multipolar polarisabilities
of the monomer generally regarded as the dominant contribution, so that 
inaccurate DFT predictions of these polarisabilities may be
problematic. In order to describe the subtle balance between these
different interaction mechanisms without adjustment
of empirical parameters, a good description of all of them
is highly desirable.

The many-body expansion of the total energy of a system of molecules
gives a helpful way of analysing the different types of interaction
in water~\cite{xantheas94,pedulla96,hankins70}, 
and has also played a key role in the construction of
interaction models fitted to molecular quantum chemistry 
calculations~\cite{ywang11}. In this expansion, 
the total energy of a system of $N$ molecules is expressed as
\begin{equation}
E = E^{(1)}+E^{(2)}+E^{(3)}+\cdots,
\label{eqn:many-body-expansion}
\end{equation}
where
\begin{eqnarray*}
   E^{(1)} &=& \sum_i E(i),\\
   E^{(2)} &=& \sum_{i<j} \delta E(i,j),\\
   E^{(3)} &=& \sum_{i<j<k} \delta E(i,j,k)\;,
\end{eqnarray*}
etc.
Throughout the paper we use the notation $E(1,2,\ldots)$ to denote the
energy of the system composed of the listed monomers, and the 2- and
3-body energy increments are given by the usual formulae:
\begin{eqnarray}
  \delta E(i,j) &=& E(i,j)- E(i) - E(j) \label{eq:2bodinc}\\
  \delta E(i,j,k) &=& E(i,j,k) - \delta E(i,j) - \delta E(i,k) -
  \delta E(j,k) - E(i) - E(j) - E(k). \notag
\end{eqnarray}
We use the notation $E^{(>2)}$ to denote all of the beyond-2-body
effects, i.e.\ $E^{(>2)} = E-E^{(1)}-E^{(2)}$.

By calculating the DMC energy for sets of configurations of H$_2$O clusters
from the monomer to the hexamer, and by comparing these energies
with many-body decompositions of both benchmark CCSD(T) energies
and of energies given by various DFT approximations, we will attempt
to probe the accuracy of DMC for the 1-, 2- and beyond-2-body parts
of the energy. The evidence to be presented will indicate that DMC is
very accurate for all the main components of the energy, while the
DFT approximations we study all encounter difficulties with one or
more of them. An important feature of this work is that many of
our configuration sets are random statistical samples
created by drawing clusters from a classical simulation of liquid
water based on an interaction model for flexible monomers. A recent
investigation of monomer and dimer energetics was based on a similar
idea~\cite{santra09}. 

Our work builds on very extensive earlier work on water
clusters in which DFT has been compared with accurate molecular
quantum chemistry 
calculations~\cite{xu04,anderson06,dahlke08,santra08,santra09,ireta04,santra07,su04},
sometimes employing the
many-body expansion, though rather little of that work
has been done on the kind of random statistical samples that 
we emphasise here. We also build on previous DMC work 
on clusters~\cite{gurtubay07,santra08}, bulk water~\cite{grossman05}
and ice structures~\cite{santra11}. The DMC work 
already reported on clusters~\cite{gurtubay07,santra08}
investigated only a single configuration of the dimer and some of
the equilibrium isomers of the hexamer, but was vital in
giving evidence for the accuracy of DMC for water systems. Very recent
work on the cohesive energy and relative energies, equilibrium
volumes and transition pressures of a number 
of ice structures~\cite{santra11}
demonstrated the remarkable accuracy of DMC ($\sim 0.2$~m$E_{\rm h} \simeq
5$~meV per monomer) in the bulk. Combining the evidence from that work
on ice with the present evidence on clusters, we shall argue that
DMC can now be regarded as a valuable tool which will be 
able to provide benchmark
energies for larger clusters, for statistical samples of configurations
of liquid water under a range of conditions, and for more complex
systems such as ice surfaces.


\section{Techniques}
\label{sec:techniques}

Quantum Monte Carlo methods have been described in detail in many
previous papers~\cite{hammond94,anderson99,nightingale99,foulkes01,needs10,towler11,austin12}.
The main technique used here is diffusion Monte
Carlo (DMC), which is a stochastic technique for obtaining the ground-state
energy of a general many-electron system. We recall that it works
by propagating an initial trial many-electron wavefunction 
in imaginary time, with the time-dependent part of the wavefunction
represented by an evolving population of walkers. Although DMC is
in principle exact, practical calculations generally employ
the fixed-node approximation~\cite{fixed-node} and the 
pseudopotential locality~\cite{pseudo-locality}
approximation. Errors arising from both these approximations are
greatly reduced by improving the accuracy of the trial wavefunction,
and we do this by variance minimisation~\cite{variance-min} 
within variational Monte Carlo. (The alternative procedure of
energy minimisation is favoured by some research groups.)
All our DMC calculations are made with the {\sc casino} 
code~\cite{casino}.

We use conventional Slater-Jastrow wavefunctions with a Jastrow factor
containing electron-nucleus, electron-electron and
electron-nucleus-electron terms, each of
which depends on variational parameters~\cite{drummond04}. We use Dirac-Fock 
pseudopotentials~\cite{dirac-fock-pseudo}, 
with the oxygen pseudopotential having
a He core with core radius $0.4$~\AA\ and the hydrogen pseudopotential
having core radius $0.26$~\AA. The single-electron orbitals
in the trial wavefunction are obtained from plane-wave LDA
calculations performed with the PWSCF code~\cite{pwscf}. 
We use LDA to generate the single-electron orbitals, because the
evidence indicates that this usually gives more accurate many-electron
trial functions than other DFT approximations or Hartree-Fock~\cite{ma09}.
These DFT calculations
are performed on water clusters in large cubic boxes having
a length of $40$~a.u. in periodic boundary conditions, with
$\Gamma$-point sampling and a plane-wave cut-off of $300$~Ry. The
resulting single-electron orbitals are then re-expanded 
in B-splines~\cite{alfe04}
and the DMC calculations are performed with free (as opposed to
periodic) boundary conditions. The B-spline grid has the natural
spacing $a = \pi / G_{\rm max}$, where $G_{\rm max}$ is the
plane-wave cut-off wavevector. 

Since DMC is a stochastic technique, its computed energies suffer from
a statistical error, which is inversely proportional to the square root
of configuration-space sampling points, so that it is desirable to increase
the number of walkers and the 
imaginary-time duration of the run. Since the recent DMC work
on ice structures~\cite{santra11} suggests that DMC is capable of an
accuracy of $0.1 - 0.2$~m$E_{\rm h}$/monomer or better for water systems,
we generally aim to run the DMC calculations long enough to reduce the
statistical error below this tolerance. 
For the imaginary-time propagation of DMC,
we use a time-step of $\delta t = 0.005$~a.u. This value was chosen after tests
with smaller $\delta t$ down to $0.001$~a.u., which showed 
that with $\delta t = 0.005$~a.u.
the total energy of the H$_2$O monomer is converged to within
$0.4$~m$E_{\rm h}$. This tolerance is outside our target of $0.1$~$E_{\rm h}$
per monomer, but we are concerned in this work with energy {\em differences},
and we expect that all the relevant differences will be converged with
respect to time-step to very much better than $0.4$~m$E_{\rm h}$
with $\delta t = 0.005$~a.u.
Our DMC results on the H$_2$O monomer (Sec.~\ref{sec:monomer}) will
confirm this. As further confirmation, we have also made tests on
the energy differences between dimer configurations with 
$\delta t$ ranging from $0.001$ to $0.005$~a.u. and we find a variation
of no more than $0.1$~m$E_{\rm h}$.

All the present DMC calculations were performed on the JaguarPF supercomputer
at Oak Ridge National Laboratory, which at the time of the calculations
consisted of 224,000 cores organised into 12-core shared-memory nodes. 
The parallel implementation of the {\sc casino} code distributes walkers
over cores. Because the walker populations fluctuate, 
we find that reasonable load-balancing is achieved only if the mean
number of walkers on each core is at least 10. The total number of
walkers and hence the optimal number of cores depend strongly
on the number of atoms in the system and the required statistical accuracy,
but in general we find it convenient to use up to about $50,000$
cores. The parallel scaling of our {\sc casino} implementation
on JaguarPF is excellent, becoming essentially perfect for very large
numbers of walkers. For the statistical samples of configurations
used in the present work, it is efficient to run a number of DMC
calculations simultaneously. 

Our absolute standard of accuracy throughout this work is the
coupled-cluster approximation CCSD(T) in the complete basis-set (CBS) limit.
Of course, this limit cannot be attained, but our convergence criterion
is that residual basis-set errors should be less than the threshold of
$0.1$~m$E_{\rm h}$/monomer mentioned above. In order to achieve this tolerance,
we use the identity
\begin{equation}
E ( {\rm CCSD(T)} ) = E ( {\rm MP2} ) + E ( \Delta {\rm CCSD(T)} ) \; ,
\end{equation}
where
$E ( \Delta {\rm CCSD(T)} ) \equiv E ( {\rm CCSD(T)} ) - E ( {\rm MP2} )$
is the change of the correlation energy when CCSD(T) is used in place
of MP2. This allows us to use different basis sets for
$E ( {\rm MP2} )$ and $E ( \Delta {\rm CCSD(T)} )$,
exploiting the fact that the size of basis set needed to converge
$E ( \Delta {\rm CCSD(T)} )$ is less than that needed to
converge $E ( {\rm CCSD(T)} )$ itself. This approach is sometimes
referred to as the ``focal point'' 
scheme~\cite{tschumper02,east93,allinger97}. All our molecular
quantum-chemistry calculations are performed with the
{\sc molpro} code~\cite{molpro,molpro:2011}.

For all the benchmark calculations, we use the Dunning
augmented correlation-consistent basis sets aug-cc-pVXZ \cite{dunning89,kendall92}, with X the
cardinality, referred to here simply as AVXZ. The basis-set
convergence of Hartree-Fock (HF) energies is rapid and unproblematic, and that
of correlation energies can be greatly accelerated by the use of F12 (explicitly correlated)
methods~\cite{kutzelnigg85,kutzelnigg91,klopper06}. 
The F12 method is available for both MP2 and CCSD(T)
in the {\sc molpro} code~\cite{werner07,adler07}, 
and we use it in all the present calculations.
In addition, the efficiency of HF and MP2 calculations, and of F12
contributions to MP2 and CCSD(T), is greatly enhanced by
using density fitting 
(DF)~\cite{werner07,manby03,werner03,polly04}, the 
errors incurred being
typically a few $\mu E_{\rm h}$, and so completely negligible for
present purposes.

Since the many-body decomposition 
(Eq.~(\ref{eqn:many-body-expansion})) plays a key role
in our analysis, we need the many-body components of the benchmark
CCSD(T) energy for every configuration studied. 
For the 1-body energy $E^{(1)}$, rather than using CCSD(T) itself,
we prefer to use the more accurate Partridge-Schwenke (PS) energy
function~\cite{partridge97}. 
This is an elaborate parameterised
function fitted to a very large number of extremely accurate
quantum chemistry calculations spanning a wide range of monomer
geometries. The function, denoted here by $E^{(1)} ( {\rm PS} )$,
has been shown to have spectroscopic accuracy, and it can therefore
be regarded as essentially exact for present purposes.
When we refer to CCSD(T) benchmarks, what we actually mean is
$E^{(1)} ( {\rm PS} ) + E^{(2)} ( {\rm CCSD(T)} ) + \ldots$.
The many-body decomposition is an additional aid to basis-set
convergence,
with smaller basis sets sufficing for high-order interaction terms.
As a further important point of technique, we systematically use the
counterpoise method \cite{boys70,helgaker00} to reduce basis-set
superposition error in all our calculations;
so for example to calculate a 2-body energy contribution
$\delta E(i,j) = E ( i, j ) - E( i ) - E( j )$
the full dimer basis sets is used for each of the three contributions. For completeness,
we note that we do not include core correlation or relativistic corrections.
These corrections were studied
for the H$_2$O dimer by Tschumper {\em et al.}~\cite{tschumper02} 
and found to be typically a few $\mu E_{\rm h}$.

To conclude our summary of techniques, we note briefly how we have
performed the DFT calculations. We use the same Dunning
correlation-consistent AVXZ basis sets that we used for the
benchmark calculations, and we find that the differences
between successive cardinalities decrease in essentially the same
way for all the functionals as in the HF calculations, so that
convergence of the total energy to our tolerance of 
$0.1$~m$E_{\rm h}$/monomer is easy to achieve. In computing the
many-body components of the energy, we use exactly the same counterpoise
methods outlined above for the benchmark calculations.


\section{DMC and DFT compared with benchmarks}
\label{sec:DMC_QC}

We will start by studying a thermal sample of configurations
of the H$_2$O monomer drawn from a classical molecular
dynamics (m.d.) simulation of
the bulk liquid based on the flexible {\sc amoeba} 
model~\cite{ren03,ren04}. The comparisons
will allow us to assess the accuracy of DMC and DFT approximations
for the 1-body energy $E^{(1)}$. For the dimer, we will 
gain insight into the 2-body energy by comparing DMC and
DFT energies with benchmarks 
for two sets of configurations:
first, the 10 stationary points which have been extensively
studied in earlier 
work~\cite{tschumper02,anderson06,huang06,huang08,smith90}; 
and second, a thermal sample of $198$ 
configurations drawn from the bulk liquid. We then
move on to comparisons for thermal samples of the trimer, tetramer,
pentamer and hexamer, which allow us to see how well the different methods
account for the beyond-2-body energy $E^{(>2)}$.
At the end of the Section,
we will present results for the many-body decomposition of the energy
of a number of stationary points of the hexamer, for which QMC
results have been reported in earlier work~\cite{santra08}.

Since all our thermal samples of cluster configurations were drawn
from the same {\sc amoeba} m.d. simulation, we summarise
the relevant details here. Monomer flexibility is one of the important
features of the {\sc amoeba} 
model~\cite{ren03,ren04}, whose parameters are adjusted to fit
selected {\em ab initio} and experimental data.
The model also accounts for many-body interactions through distributed
polarisabilities of the monomers.
It is known to give a rather good description of the
radial distribution functions and the self-diffusion coefficient
of liquid water over quite a range of conditions, including ambient.
The m.d. simulation run from which we drew the configurations was
performed on a system of 216 water molecules in periodic
boundary conditions at the ambient density of $1.0$~gm/cm$^3$
and room temperature (300~K). 
This way of making thermal samples is motivated by our long-term aim of
obtaining accurate descriptions of the energetics of condensed phases of
water in thermal equilibrium. To this end, it is natural
to demand that the methods we use should be accurate for cluster
geometries typical of those found in the condensed phases of interest.
Naturally, we have to
bear in mind always that the mean and rms errors we find on
comparing a chosen technique with benchmark energies for a thermal
sample are not absolute quantities, but will depend on the way 
the sample was constructed. 

\subsection{The monomer}
\label{sec:monomer}

All our calculations on the monomer were performed on a set of
100 configurations drawn at random from the {\sc amoeba}
m.d. simulation. The mean value of the O-H bond length in this
sample was 0.968~\AA, and the probability distribution of bond-lengths
was roughly Gaussian, with an rms fluctuation about the mean
of 0.019~\AA; the minimum and maximum O-H bond lengths occurring in
the sample were 0.913 and 1.020~\AA. The mean and rms fluctuation
of the H-O-H bond angle were $105.3^\circ$ and $3.7^\circ$
respectively, with the minimum and maximum angles being 
97.0$^\circ$ and 117.0$^\circ$.

We assess the errors of DMC and DFT approximations for the monomer
by comparing their energies with the essentially exact Partridge-Schwenke
energy function $E^{(1)} ( {\rm PS} )$ referred 
to in Sec.~\ref{sec:techniques}.
The equilibrium monomer geometry according to
PS has bond lengths of $0.95865$~\AA\ and 
a bond angle of $104.348^\circ$. It is
convenient to take this as the ``standard'' geometry of the
isolated monomer, so that when we refer to the energy of the monomer
in any geometry computed with a given method we will always mean
the energy of that geometry minus the energy in the PS equilibrium
geometry computed with the same method. This means that the monomer
energy with PS itself is by definition non-negative, but with a DFT
approximation or with DMC it can be negative, since the minimum-energy
configurations with these methods will generally differ from the
PS equilibrium geometry.

\begin{figure}[!htb]
\centerline{
\includegraphics[width=0.8\linewidth]{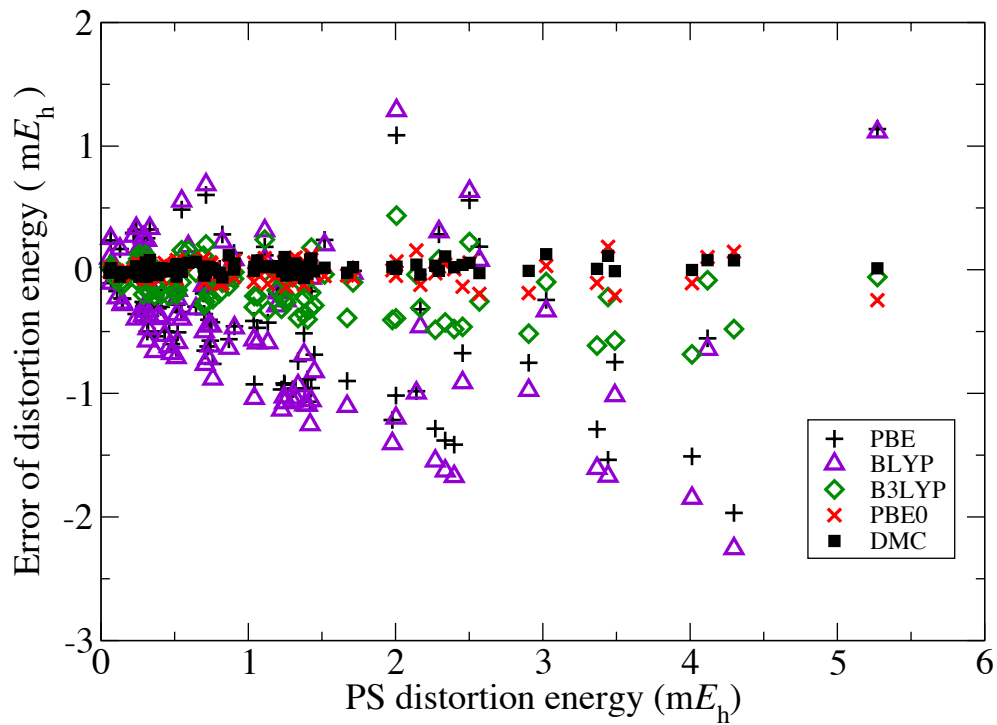}
}
\caption{Errors of DFT and DMC distortion energy of the H$_2$O monomer
for a thermal sample of 100 configurations (see text). Quantities
shown are deviations of calculated energies from Partridge-Schwenke
(PS) benchmark values with PBE (black pluses), BLYP (purple triangles), B3LYP (green diamonds)
and PBE0 (red crosses) and with DMC (black squares) plotted against the
PS distortion energy itself. Units: m$E_{\rm h}$}
\label{fig:monomer_DMC-ps_DFT-ps}
\end{figure}

Our DMC calculations on the 100 monomer configurations were performed
as described in Sec.~\ref{sec:techniques}. In our production results,
the Jastrow factor was not re-optimised for each geometry.
We used a smaller set of 25 configurations to test the effect of 
re-optimising it, and we could not detect any significant
changes of energy due to re-optimisation.
For every geometry, the DMC calculations
were continued long enough to reduce the rms statistical errors
to $30$~$\mu E_{\rm h}$. We show in Fig.~\ref{fig:monomer_DMC-ps_DFT-ps} 
$E(\text{DMC})-E(\text{PS})$
plotted against $E(\text{PS})$. The distortion energy itself
covers a range up to $\sim 5$~m$E_{\rm h}$ ($140$~meV), and we see
that DMC errors are only a tiny fraction of this. In fact, for the given
thermal sample, the mean DMC error and its rms fluctuation about the mean
are $10$ and $40$~$\mu E_{\rm h}$ (Table~\ref{tab:monomer_energy}). 

We have made comparisons against PS also for DFT, using the
PBE, BLYP, B3LYP and PBE0 exchange-correlation functionals and the
AV5Z basis. Our convergence tests using AVQZ indicate mean (rms) errors due to
basis-set incompleteness of 
around 4 (10) $\mu E_{\rm h}$ for all functionals, suggesting that any residual
basis-set errors beyond AV5Z will be much smaller than $10$~$\mu E_{\rm h}$
($0.27$~meV).

The differences $E(\text{DFT})-E(\text{PS})$ are also displayed 
in Fig.~\ref{fig:monomer_DMC-ps_DFT-ps}, 
and the mean and rms fluctuations are recorded in Table~\ref{tab:monomer_energy}. 
We see from this that PBE and
BLYP give very poor results, and their negative deviations from the 
PS benchmarks imply that the energy needed to distort the
monomer is considerably underestimated by both approximations, as already
noted in a recent paper by 
Santra {\em et al.}~\cite{santra09}. The B3LYP functional
is much better, but PBE0 is the clear winner out of the functionals
examined, with mean and rms errors of 
only $-10$ and $80$~$\mu E_{\rm h}$. The good performance
of PBE0 for the monomer was also reported by 
Santra {\em et al.}~\cite{santra09}.
However, DMC is markedly superior, and it appears 
that residual DMC errors in the 1-body energy can safely be
neglected in cluster and bulk systems, at least so long as the
variation of the bond lengths and bond angle are not too much
larger than those treated here.

\subsection{The dimer}
\label{sec:dimer}

\subsubsection{The stationary points}
\label{sec:stationary}

The geometries of the 10 dimer stationary points are depicted in many previous
papers (see e.g. Ref.~\cite{tschumper02}), 
and there is a standard numbering, which we follow here.
We have worked with two closely related sets of configurations
for the stationary points. The configurations of Tschumper
{\em et al.}~\cite{tschumper02}
were used for the tests we performed to check that our techniques
can deliver dimer energies with CCSD(T) within 
$100$~$\mu E_{\rm h}$ of the CBS limit. However, because of the
way the project developed, the DMC and DFT calculations were performed
on a slightly different configuration set due to the 
Bowman group~\cite{huang08}, and
we used exactly the same techniques tested on the Tschumper set
to produce our CCSD(T) benchmarks for the Bowman set.

Our benchmarks for the total dimer energy are represented as 
\begin{eqnarray}
  E = E^{(1)}(\mathrm{PS}) 
+ E^{(2)}(\text{MP2-F12/AV5Z}) 
+ E^{(2)}(\Delta\text{CCSD(T)-F12/AVQZ}) 
\label{eq:E1E2dimer}
\end{eqnarray}
with full counterpoise for all energies. Our basis-set tests
show that with AVQZ the HF and MP2-F12 correlation energies have residual basis-set errors of
less than $\sim 15$ and $\sim 10$~$\mu E_{\rm h}$ respectively, while with AVTZ
the basis-set errors in $E^{(2)} ( \Delta {\rm CCSD(T)-F12} )$
are less than $\sim 30$~$\mu E_{\rm h}$. For the present calculations
on the stationary points, we have gone beyond this so that the
remaining basis-set errors should be much less than those just quoted,
and our comparisons with the energies reported by Tschumper {\em et al.}
confirm this.

Our DMC calculations on the total energies of the stationary points
are computed as described in Sec.~\ref{sec:techniques}.
The runs were continued long enough to reduce the
DMC statistical error to $77$~$\mu E_{\rm h}$. Since we focus here
on the energies of the stationary points relative to that of the global
minimum, we have extended the run on the global minimum so that its
rms statistical error is only $44$~$\mu E_{\rm h}$. The relative DMC energies
are compared with the CCSD(T) benchmarks in 
Table~\ref{tab:sp_bench_DMC_DFT}, and we see that
they agree in all but two cases within better than $100$~$\mu E_{\rm h}$
($2.7$~meV), and in those two cases the DMC errors are still less than
170~$\mu E_{\rm h}$.

We already know from earlier work~\cite{anderson06} that DFT predictions of the
energies of the stationary points suffer from much larger
errors than the DMC errors just mentioned, but we considered it worthwhile
to calculate our own values using PBE, BLYP, B3LYP and PBE0.
(We note that the present DFT energies are all calculated
at exactly the same geometries, rather than at the stationary-point
geometries that would be given by the DFTs themselves, and this should
be borne in mind when comparing with earlier DFT results.)
Our tests of basis-set convergence for DFT 
show that with AV5Z the relative energies of the stationary
points are converged to better than $20$~$\mu E_{\rm h}$. We report
values of these relative energies for the Bowman geometries in
Table~\ref{tab:sp_bench_DMC_DFT}. We see that the DFT approximations
overestimate all the energy differences between the stationary points
and the global minimum, with many of the DFT errors being greater
than $0.5$~m$E_{\rm h}$ and a few being as much as $1.0$~m$E_{\rm h}$,
in general agreement with earlier work~\cite{anderson06}.
The hybrid functionals B3LYP and PBE0 perform slightly better than
non-hybrid PBE and BLYP, but there is not a great deal to choose between
them. The individual monomers
have almost exactly the same geometries at the 10 stationary
points, so that DFT errors in the monomer distortion energies play almost
no role here, and any disagreements with the benchmarks are due almost
entirely to the 2-body energies.

We conclude from these comparisons that DMC gives better (in most cases,
much better) agreement with the benchmarks than any of the DFT
approximations we have studied, with DMC errors being typically
five times smaller than than those of DFT. 

\subsubsection{A thermal sample of dimer configurations}
\label{sec:dimer_thermal}

The thermal sample was produced by drawing 198 configurations
at random from the {\sc amoeba} simulation, with O-O distances
included out to $7.5$~\AA\ but with a bias towards shorter distances.
To construct the CCSD(T) benchmark energies for this set, we followed
the procedure outlined above for the stationary points. 
The benchmark energy is computed as 
\begin{eqnarray*}
 E &=& E^{(1)}(\text{PS}) + 
 E^{(2)}(\text{MP2-F12/AVQZ})
+ E^{(2)}(\Delta\text{CCSD(T)-F12/AVTZ}).
\end{eqnarray*}
Because the configuration sample is reasonably large, we
are able to analyse the statistics of basis-set differences for the
three components. It turns out that the differences depend rather
uniformly on $R_{\rm O O}$, providing a simple scheme for partially correcting
residual basis-set errors in the MP2-F12 and $\Delta$CCSD(T)-F12
correlation contributions.  Tests
indicate that our benchmark dimer energies are
within $\sim 20$~$\mu E_{\rm h}$ of the CBS limit of CCSD(T).

\begin{figure}[!htb]
\centerline{
\includegraphics[width=0.8\linewidth]{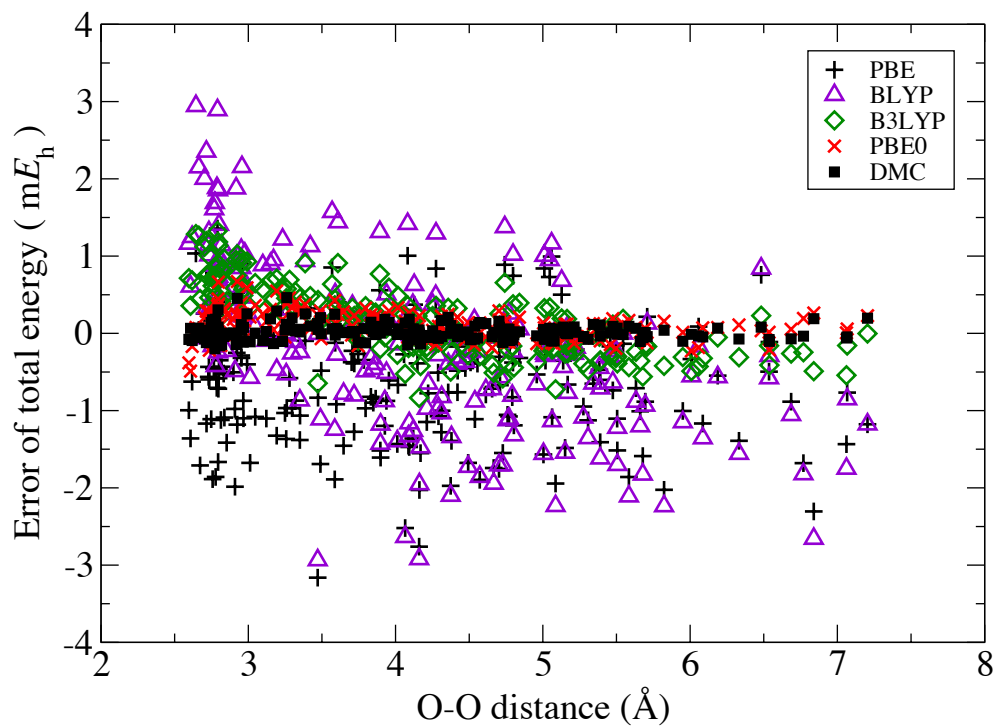}
}
\caption{Errors of DMC and DFT approximations relative to CCSD(T)
benchmarks for total energies of thermal sample of 198 dimer configuration,
plotted vs O-O distance. Symbols represent PBE (black pluses),
BLYP (purple triangles), B3LYP (green diamonds) and PBE0 (red crosses)
and DMC (black squares). Units: m$E_{\rm h}$.} 
\label{fig:dimer_tot_errors}
\end{figure}

The DMC calculations were performed in the same way as for
the stationary points, with the single-electron orbitals and Jastrow
factor constructed as described in Sec.~\ref{sec:techniques} and the time-step
chosen as before to have the value $0.005$~a.u. For every configuration,
the DMC run was continued long enough to reduce the statistical
error on the total energy to $60$~$\mu E_{\rm h}$.
To characterise the performance of DMC, we show in 
Fig.~\ref{fig:dimer_tot_errors} the energy
difference DMC-bench plotted as a function of $R_{\rm O O}$. The
differences are typically on the order of $100$~$\mu E_{\rm h}$, and
there is no obvious dependence of their magnitude on $R_{\rm O O}$.
Quantitatively, the mean value of $E ( {\rm DMC} ) - E ( {\rm bench} )$
over the 198-configuration sample is $20$~$\mu E_{\rm h}$ and the
rms fluctuation is $0.10$~m$E_{\rm h}$. Since the statistical
errors of the Monte Carlo sampling in our DMC calculations are 
$\sim 60$~$\mu E_{\rm h}$, this means that there are
statistically significant deviations of the DMC
energies from the CCSD(T) benchmarks, but
they appear to be no more than $\sim 0.1$~m$E_{\rm h}$ ($2.7$~meV).

Our DFT calculations on the thermal sample were done both as direct
calculations of the total energy using AV5Z basis set, and also
by separating the total energy into 1- and 2-body
parts, using the AV5Z basis for $E^{(1)}$ and AVQZ with counterpoise
for $E^{(2)}$. A comparison of the direct total energies calculated
using AVQZ and AV5Z basis sets shows mean and rms differences of
the energies that are less than $25$~$\mu E_{\rm h}$ for all the
functionals. The total energies obtained in the direct calculations
differ from those obtained by adding the 1- and 2-body energies
by at most $50$~$\mu E_{\rm h}$.
The differences $E ( {\rm DFT} ) - E ( {\rm bench} )$
for the dimer total energies are plotted against $R_{\rm O O}$
in Fig.~\ref{fig:dimer_tot_errors}. 
It is immediately clear that the DFT errors are considerably
greater than those of DMC, with BLYP and PBE being much inferior to
B3LYP and PBE0. The mean values of the DFT-bench differences
and their rms fluctuations about these means are reported 
in Table~\ref{tab:dimer_errors}.

\begin{figure}[!htb]
\centerline{
\includegraphics[width=0.8\linewidth]{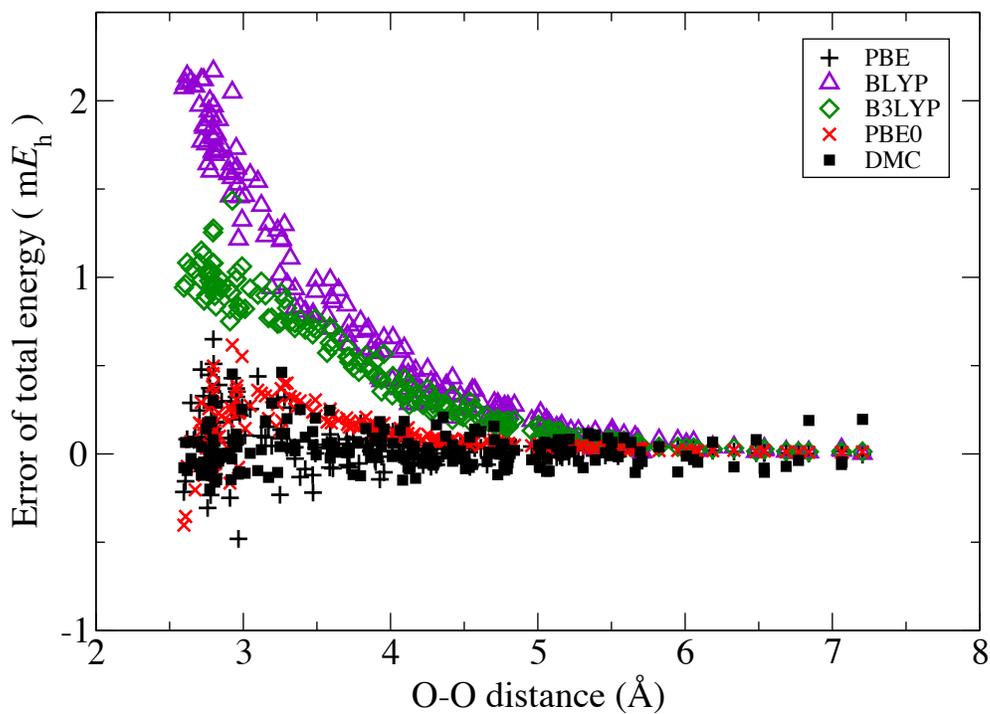}
}
\caption{Errors of DFT approximations for total energies
of thermal sample of 198 dimer configurations when 1-body part is corrected 
by replacing the DFT 1-body energy by the essentially exact 
Partridge-Schwenke function. As in Fig.~\ref{fig:dimer_tot_errors},
errors are relative to CCSD(T) benchmarks and are plotted vs
O-O distance. Symbols represent PBE (black pluses), BLYP (purple triangles),
B3LYP (green diamonds) and PBE0 (red crosses).
Errors of DMC (black squares) are shown for comparison. Units: m$E_{\rm h}$.
}
\label{fig:dimer_2b_errors}
\end{figure}

An important theme of the present work is the separation of the energy into
its many-body parts. We have already seen that some of the DFT approximations
suffer from large 1-body errors, so it is natural to ask how they perform when
these errors are corrected. If we separate the total energy $E({\rm DFT})$ 
computed with a given DFT into its 1-body and 2-body parts, and we then
replace the 1-body energy by its Partridge-Schwenke value
$E^{(1)} ( {\rm PS} )$, we obtain an approximation denoted here
by DFT-$\Delta_1$. Clearly, for dimers the errors of DFT-$\Delta_1$
are entirely 2-body errors.  In Fig.~\ref{fig:dimer_2b_errors}, we compare
the total-energy differences $E ( {\rm DFT} \mbox{-} \Delta_1 ) - E ( {\rm bench} )$ as a function
of $R_{\rm OO}$ with the differences $E ( {\rm DMC} ) - E ( {\rm bench} )$. 
The mean and rms values of
these differences are reported in Table~\ref{tab:dimer_errors}. We see that BLYP is very poor
indeed, being much too repulsive over the whole range $2.5 - 4.5$~\AA, and
B3LYP suffers from the same problem, though its errors are smaller. The
PBE0 approximation is much better, though it still too repulsive. 
Best of all these DFTs is PBE. The DMC errors are even small than those of PBE.

These findings are generally in line with what is known from previous
DFT and DMC work on the binding energy of the dimer. It is well
known that BLYP seriously underbinds, that B3LYP underbinds somewhat less,
and that PBE0 and PBE give good binding energies, with PBE being almost
exactly correct. It is also known that DMC gives a rather accurate value of
the dimer binding energy~\cite{gurtubay07}. 
The present results enlarge the picture by showing
that these errors in the binding energy at the global minimum can be
seen as part of general trends over a range of O-O distances.
 
\subsection{The trimer}
\label{sec:trimer}

The trimer is important, because it is the smallest cluster for which
we can probe beyond-2-body interactions. 
We created a thermal sample of trimer configurations
using the same {\sc amoeba} simulation of liquid water as before,
drawing 50 trimer geometries at random, with the condition that all three
O-O distances must be less than $4.5$~\AA. 

Our method for obtaining basis-set converged CCSD(T) energies is
a straightforward extension of what we outline above for the dimer.
The total trimer energy is decomposed as 
\begin{eqnarray*}
  E 
    &=& 
  E^{(1)} + E^{(2)} +
  E^{(3)}(\text{MP2-F12/AVQZ}) +
  E^{(3)}(\Delta\text{CCSD(T)-F12/AVTZ}) 
\label{eq:E3trimer}
\end{eqnarray*}
where the terms $E^{(1)} + E^{(2)}$ are treated exactly as in Eq.~\ref{eq:E1E2dimer}. We find that 
the Hartree-Fock component $E^{(3)} ( {\rm HF} )$
converges very rapidly with basis set: the mean difference
and the rms fluctuation for AVQZ $-$ AVTZ are only
$1.0$ and $1.1$~$\mu E_{\rm h}$. The same is true for 
the correlation part of $E^{(3)} (\text{MP2-F12} )$, for which the corresponding values
are $1.0$ and $1.6$~$\mu E_{\rm h}$. For 
$E^{(3)} ( \Delta \text{CCSD(T)-F12} )$,
we have results only for AVTZ. However, our calculations on the
dimer samples showed that the AVQZ $-$ AVTZ difference for
CCSD(T)-F12 was very similar to that of MP2-F12, and we assume the same to be true
for $E^{(3)}$. 
From the evidence we have obtained, the energy expression in Eq.~\ref{eq:E3trimer} should
be well within $50$~$\mu E_{\rm h}$ of CCSD(T)/CBS.

\begin{figure}[!htb]
\centerline{
\includegraphics[width=0.8\linewidth]{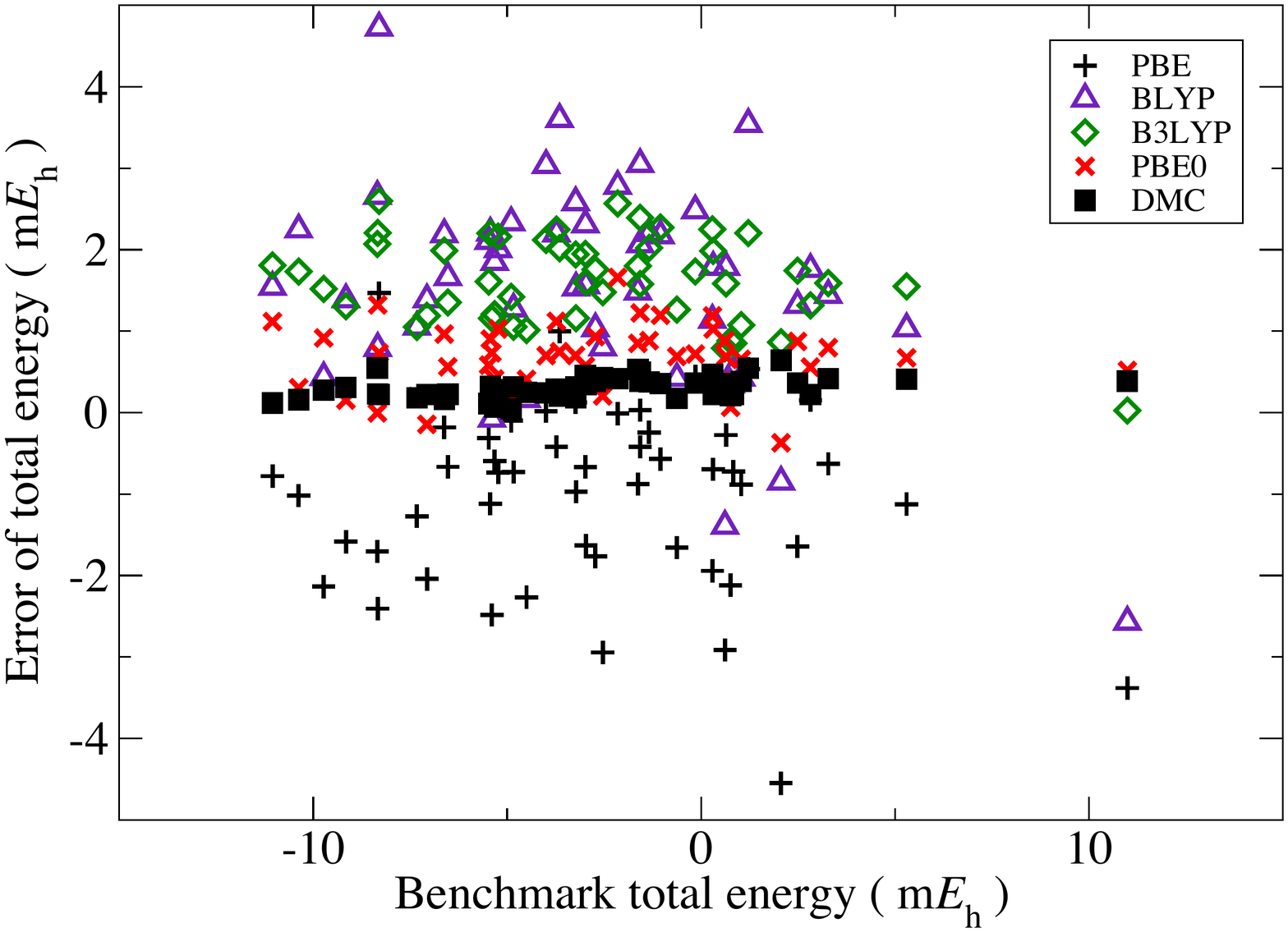}}
\caption{Errors of DFT and DMC total energy of the H$_2$O trimer
for a thermal sample of 50 configurations drawn from a classical
simulation of liquid water (see text). Quantities shown are deviations
of calculated energies from CCSD(T) benchmark energies near the
basis-set limit, with PBE (black pluses), BLYP (purple triangles),
B3LYP (green diamonds) and PBE0 (red crosses) and with
DMC (black squares) plotted against the benchmark energy itself.
Units: m$E_{\rm h}$}
\label{fig:tr_errors_DMC_DFT}
\end{figure}

Our DMC calculations on the 50-configuration trimer sample follow
exactly the methods outlined in Sec.~\ref{sec:techniques}. The duration
of the DMC runs was long enough to reduce the statistical error
on the total energy to $77$~$\mu E_{\rm h}$.
We show in Fig.~\ref{fig:tr_errors_DMC_DFT} 
the errors $E ( {\rm DMC} ) - E ( {\rm bench} )$
of the DMC total energy of the trimers 
plotted against the benchmark total energy. We see
that the errors and their fluctuations are very small, their mean value
and rms fluctuation over the 50-configuration set being
$0.30$~m$E_{\rm h}$ (8.1~meV) and $0.13$~m$E_{\rm h}$ ($3.5$~meV). (Note
that these values refer to the {\em total} energy, not the
energy per monomer.)

DFT calculations of the total trimer energy are straightforward,
and our tests indicate that the total energy relative to that of three
monomers in the PS reference geometry is converged to within
$0.15$~m$E_{\rm h}$ with AVQZ basis sets. It is useful to have the
many-body decomposition, and we use 
$$
E ( {\rm DFT} ) = E^{(1)} ( {\rm DFT/AV5Z} ) + E^{(2)} ( {\rm DFT/AVQZ} ) +
E^{(3)} ( {\rm DFT/AVTZ} ),
$$
again with full counterpoise correction.
The two ways of calculating the total trimer energies agree to within
mean and rms differences of $120$ and $56$~$\mu E_{\rm h}$ respectively. 
As we show in Fig.~\ref{fig:tr_errors_DMC_DFT}, the errors
of the DFT total energy with PBE,
BLYP, B3LYP and PBE0 are much greater than those of DMC:
the mean and rms deviations (Table~\ref{tab:compare_DFT-n}) are typically
five times greater than the DMC values.

It is now interesting to analyse how much of the DFT errors come from
1-, 2- and 3-body parts of the energy. Here we follow the approach of
Taylor \emph{et al.} \cite{taylor12}, correcting the low-order
many-body contributions to DFT.
For example, the effect of 1-body errors can be eliminated by using
the energy expression
\begin{eqnarray*}
E ( \text{DFT-}\deltaone ) &=& E ( \text{DFT} ) +
E^{(1)} ( \text{bench} ) - E^{(1)} ( \text{DFT} )  \\
&\equiv&
E^{(1)} ( \text{bench} ) + E^{(2)} ( \text{DFT} ) + E^{(3)} ( \text{DFT} ).
\end{eqnarray*}

The mean and rms deviations
$E ( \text{DFT-}\deltaone ) - E ( \text{bench} ) =
[ E^{(2)} ( \text{DFT} ) - E^{(2)} ( \text{bench} ) ] +
[ E^{(3)} ( \text{DFT} ) - E^{(3)} ( \text{bench} ) ]$ are reported
in Table~\ref{tab:compare_DFT-n}. As 
expected from the results of Sec.~\ref{sec:monomer}, the rms
fluctuations of $E ( \text{DFT-\deltaone} ) - E ( \text{bench} )$
are very much reduced for PBE and BLYP, because their 1-body
energies are poor; the mean value of the deviation is improved for PBE
but worsened for BLYP, again as expected. On the other hand for B3LYP
and particularly for PBE0, correction of the 1-body errors makes little
difference.

We can go further by correcting both the 1-body and the 2-body
DFT energies, thus obtaining a scheme that we denote by DFT-\deltatwo.
The trimer energy in this scheme is $E ( \text{DFT-\deltatwo} ) =
E ( \text{DFT-\deltaone} ) + [ E^{(2)} ( \text{bench} ) - E^{(2)} (
\text{DFT} ) ]$,
which is the same as $E^{(1)} ( \text{bench} ) +
E^{(2)} ( \text{bench} ) + E^{(3)} ( \text{DFT} )$. The deviations
$E ( \text{DFT-\deltatwo} ) - E ( \text{bench} ) =
E^{(3)} ( \text{DFT} ) - E^{(3)} ( \text{bench} )$ now come entirely
from DFT errors in the 3-body part. The mean and rms values
of these deviations are reported in Table~\ref{tab:compare_DFT-n}. 
Not surprisingly, these
values are very small, so that 3-body effects are quite well represented
by all the DFT approximations.

It is clear that the trimer is really too small to yield very 
interesting conclusions about the accuracy of DFT compared with DMC
for beyond-2-body interactions, because there is only a single
3-body term in the total energy. However, as we go to larger clusters,
the number of beyond-2-body interactions increases rapidly, so that more
interesting comparisons can be made. We therefore turn next to
benchmark, DMC and DFT calculations on the tetramer, pentamer and hexamer.

\subsection{Thermal samples of the tetramer, pentamer and hexamer}
\label{sec:tet_pent_hex}

Our procedures for the tetramer, pentamer and hexamer follow quite
closely those outlined above for the smaller clusters. To generate the
samples for cluster size $N \ge 4$, we repeatedly planted a sphere of
chosen radius $R ( N )$ at a random position at random time-steps
of the {\sc amoeba} simulations, and if the number of molecules inside
the sphere was equal to $N$ we accepted these molecules as a sample
configuration. (For this purpose, a molecule was counted as being inside
the sphere if the O position was inside the sphere.) In making the sample
of configurations for a given $N$, we chose $R ( N )$ so that the mean
number of molecules found inside the sphere was close to $N$. In this way,
we formed samples of 25 configurations each for $N = 4$, $5$ and $6$.

Our benchmark energies were computed as
\begin{eqnarray*}
  E 
    &=& 
  E^{(1)}(\text{PS}) +
  E^{(2)}(\text{MP2-F12/AV5Z}) +
  E^{(2)}(\Delta\text{CCSD(T)-F12/AVTZ}) \\
    &+&
  E^{(3)}(\text{MP2-F12/AVTZ}) +
  E^{(3)}(\Delta\text{CCSD(T)-F12/AVDZ}) \\&+&
  E^{(>3)}(\text{MP2-F12/AVDZ}) 
\label{eq:Egeneral}
\end{eqnarray*}
for the tetramer and pentamer, but for the hexamer we use
$E^{(2)}(\text{MP2-F12/AVQZ})$ in place of
$E^{(2)}(\text{MP2-F12/AV5Z})$.
Contributions up to 3-body were treated using the counterpoise
correction as described above; higher-order terms were computed using
full counterpoise for the entire cluster;
for example, in computing 4-, 5- and 6-body
terms for the hexamer, we used the full basis set of the entire cluster
for every contribution. 

\begin{figure}[!htb]
\centerline{
\includegraphics[width=0.8\linewidth]{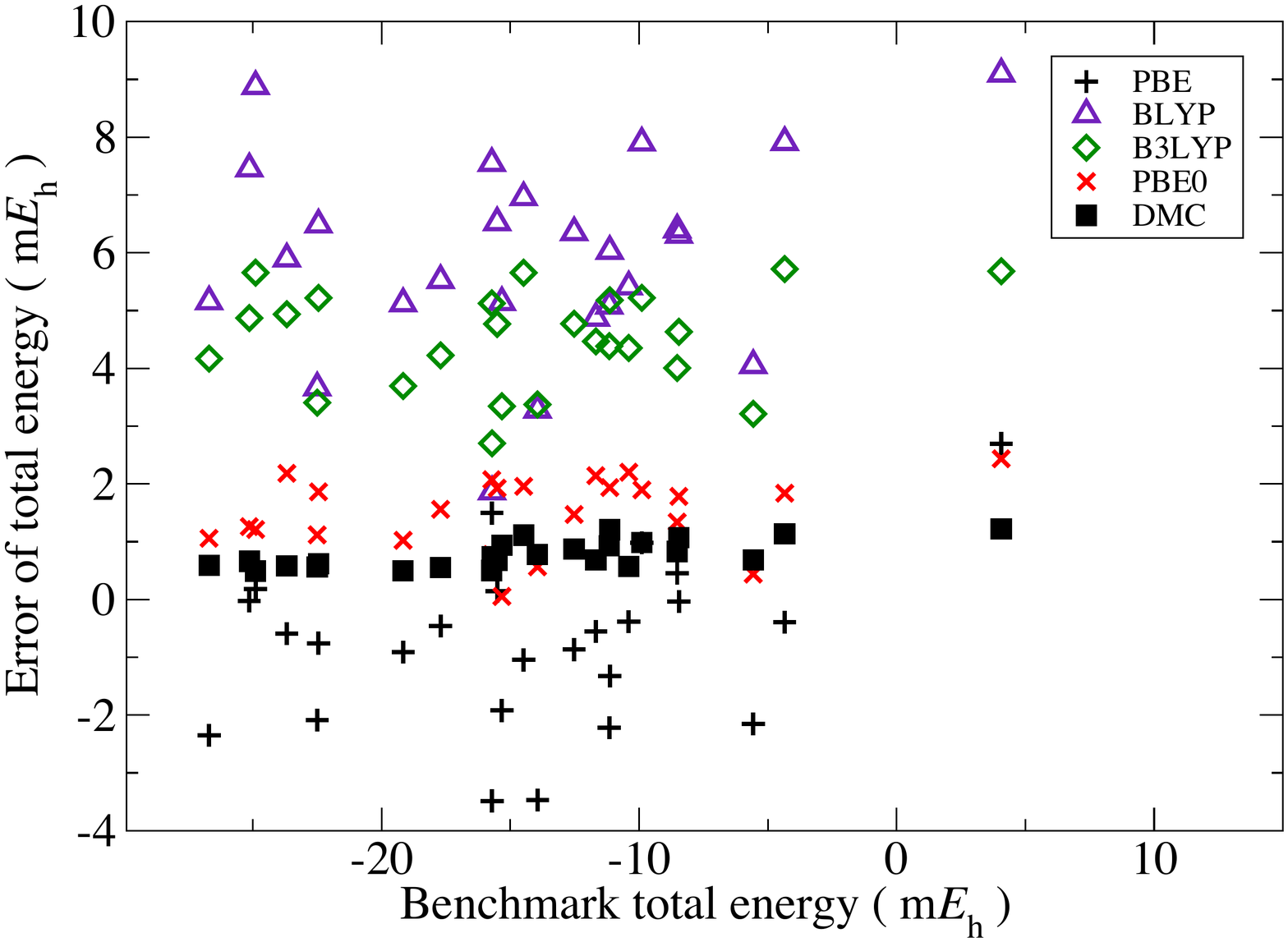}}
\caption{Errors of DFT and DMC total energy of the H$_2$O pentamer
for a thermal sample of 25 configurations drawn from a classical
simulation of liquid water (see text). Quantities shown are deviations
of calculated energies from CCSD(T) benchmark energies near the
basis-set limit, with PBE (black pluses), BLYP (purple triangles),
B3LYP (green diamonds) and PBE0 (red crosses) and with
DMC (black squares) plotted against the benchmark energy itself.
Units: m$E_{\rm h}$}
\label{fig:pe_errors_DMC_DFT}
\end{figure}

The DMC calculations for the 25-configuration samples of the
tetramer, pentamer and hexamer followed exactly the same
procedures as before, and the runs were continued until the
statistical errors on the total energy were reduced to
$0.13$, $0.14$ and $0.17$~m$E_{\rm h}$ for the tetramer, pentamer
and hexamer respectively. As an example of the very
close agreement between DMC and the CCSD(T) benchmarks, we show in
Fig.~\ref{fig:pe_errors_DMC_DFT}
the total-energy difference DMC-benchmark for the
pentamer configurations plotted against the total energy itself.
The mean value and the rms fluctuations of this difference are reported
for all the clusters in Table~\ref{tab:compare_DFT-n}. 
We see from this that the DMC
errors are on the same scale of $\sim 0.2$~m$E_{\rm h}$/monomer or less
that characterise the recently reported DMC values of the absolute and
relative energies of various ice structures~\cite{santra11}.

The DFT energies of the $N \ge 4$ cluster configurations were all
computed as sums of many-body contributions. For the 1-, 2- and 3-body
energies, we employed AV5Z, AVQZ and AVTZ basis sets respectively,
with full counterpoise for all dimers
and trimers for the 2- and 3-body energies. In the $(n \ge 4)$-body
parts, we found it more convenient to use full counterpoise for the
entire cluster, as we did for the CCSD(T) benchmarks, and for this
purpose we used AVDZ basis sets. To cross-check the total energies
obtained in this way, we also computed them directly (i.e. without
many-body decomposition), using AVQZ basis sets.

As an example of DFT performance, we show 
in Fig.~\ref{fig:pe_errors_DMC_DFT} the differences
DFT-benchmark for the pentamer sample plotted against the benchmark
total energies. The mean values of these differences and their
rms fluctuations are reported for all the larger clusters 
in Table~\ref{tab:compare_DFT-n}.
It is evident that the accuracy of DMC is very much greater
than any of the DFT approximations. As might be expected from our
results for the smaller clusters, BLYP is very poor, having rms deviations
from the benchmarks that are about 10 times the size of those with DMC,
and its mean deviations are also large. For all the clusters, PBE is
somewhat better and B3LYP still better, but best of all is
PBE0, whose rms errors are a little over 2.5 times those of DMC.

In our discussion of DFT errors for the 
trimer (Sec.~\ref{sec:trimer}), we pointed
out the possibility of correcting DFT first for 1-body errors and then
for both 1-body and 2-body errors, these two levels of corrected DFT
being denoted by DFT-\deltaone\ and DFT-\deltatwo. Since we have all the $n$-body
parts of both benchmark and DFT energies of the tetramer, pentamer
and hexamer, we can make the same analysis for them. We report in
Table~\ref{tab:compare_DFT-n} the mean and rms 
deviations of the DFT-\deltaone\ and DFT-\deltatwo\ energies 
away from the benchmarks. As expected, correction of the 1-body
energy substantially reduces the rms errors of PBE and BLYP, because these
DFTs have quite large 1-body errors, but it makes rather little
difference in the case of B3LYP and PBE0, because their 1-body errors
are small. Interestingly,
this correction considerably worsens the mean BLYP errors, because
in the uncorrected version there is a partial cancellation of
errors between the 1- and 2-body parts. Correcting for both 1- and 2-body
errors, the approximations suffer only from $( n \ge 3)$-body errors, which
we also report in Table~\ref{tab:compare_DFT-n}. Clearly, 
these errors are extremely small.
Indeed, the corrected DFTs B3LYP-\deltatwo and PBE0-\deltatwo are even slightly
better than DMC.

The comparisons we have presented demonstrate the high accuracy of DMC,
but they also indicate that the 1-body, 2-body and beyond-2-body
parts of the total energy are individually well described by DMC.
However, there is another aspect of DMC predictions that is worth
examining. If we judged solely by our comparisons for the thermal
samples, we would infer that all the main errors of DFT are in the 1-
and 2-body parts, so that once these are corrected we get approximations
that are as good as DMC. However, this inference is not true in general,
as we will show next by a many-body analysis of the stable
isomers of the hexamer.

\subsection{Stable isomers of the hexamer}
\label{sec:hexamer}

The global- and local-minimum structures of the H$_2$O hexamer
have long played a role in the understanding of
water energetics, because their relative energies are determined
by a rather delicate interplay between different kinds of 
interactions~\cite{olson07,dahlke08,santra08,bates09,kumar10,kim94b,liu96,xantheas02}.
The isomers we will be concerned with here are the prism, cage, book
and ring, whose geometries have been presented in many previous 
papers (see e.g. Ref.~\cite{santra08}).
The atomic coordinates that we use here are the MP2/AVTZ-optimised
structures of Santra {\em et al.}~\cite{santra08}.
The more open structures, such as the ring, favour hydrogen
bonding with OH$\cdots$O angles that maximise the strengths of individual
hydrogen bonds. In the more compact structures, including the prism and
cage, the total number of hydrogen bonds is greater, but the OH$\cdots$O
angles are less favourable. The book structure is a compromise betweeen
the two kinds. Coupled-cluster CCSD(T) calculations near the basis-set
limit leave no doubt that the more compact structures are more stable,
the consensus being that the prism is the global minimum, with the cage
slightly above it~\cite{bates09}. The ring is 
less stable by $\sim 3.0$~m$E_{\rm h}$
in the total energy, and the book has an intermediate energy. DMC
calculations give the correct energy ordering and energy differences
that agree closely with the CCSD(T) values~\cite{santra08}, and we noted in the
Introduction that this is one of the key pieces of evidence for the
accuracy of DMC. Most of the conventional DFT approximations give the wrong
energy ordering, with BLYP and B3LYP making the ring the global minimum, and
PBE giving this honour to the book~\cite{dahlke08}. The 
reasons for this have been
widely discussed, and a many-body analysis has already been used
to identify the cause of the problems, the suggestion being
that the lack of long-range dispersion 
is responsible~\cite{santra08}. A detailed break-down of the contributions to the relative
energies of isomers of the hexamer has also been reported 
by Wang {\em et al.}~\cite{wang10}. We find it
worthwhile to revisit this question, because DMC can now be compared with
more accurate CCSD(T) results than were 
available before and because
our own many-body analysis
indicates that the lack of dispersion may not be the only cause
of DFT errors.

We report in Table~\ref{tab:hex_isomer} our MP2 
and CCSD(T) results for the prism, cage,
book and ring isomers, computed in all cases with the MP2-optimized
structures given by 
Santra {\em et al.}~\cite{santra08}, which are also the structures
used in their DMC calculations. Our MP2 energies come from direct
calculations on the entire hexamer, using AVQZ basis sets with F12.
The CCSD(T) energies reported in the Table,
do not, however, come from direct calculations on the cluster; 
instead, we compute
$$
 E = E(\text{MP2-F12/AVQZ}) + \sum_{i=1}^3E^{(i)}(\Delta\text{CCSD(T)-F12})
$$
As shown in the Table, 
our MP2 energy differences
between the isomers agree very closely with earlier
highly converged results,
and our CCSD(T) energy differences
also agree very well with recent CCSD(T) results close to the CBS limit. 
In agreement with previous work, we find
that on going from MP2 to CCSD(T) the energy difference between the prism
and the cage increases significantly, and the energy of the ring
above the prism increases by $\sim 1.0$~m$E_{\rm h}$. The Table also
gives the DMC energy differences of Santra {\em et al.}, and we note
that they agree very closely with the CCSD(T) results reported here.
Importantly, DMC is closer to the CCSD(T) energies than to those
from MP2.  Our own PBE, BLYP, B3LYP and PBE0 energies computed
with AV5Z basis sets
are also included in the Table. We fully confirm the conclusions
from previous work that these DFTs give completely erroneous trends,
wrongly predicting that the less compact isomers are more stable than
the compact ones. The quantitative errors of the DFT energy differences
are substantial, with the energy difference between the ring and the prism
being in error by as much as $5$~m$E_{\rm h}$ in some cases.

\begin{figure}[!htb]
\centerline{
\includegraphics[width=0.8\linewidth]{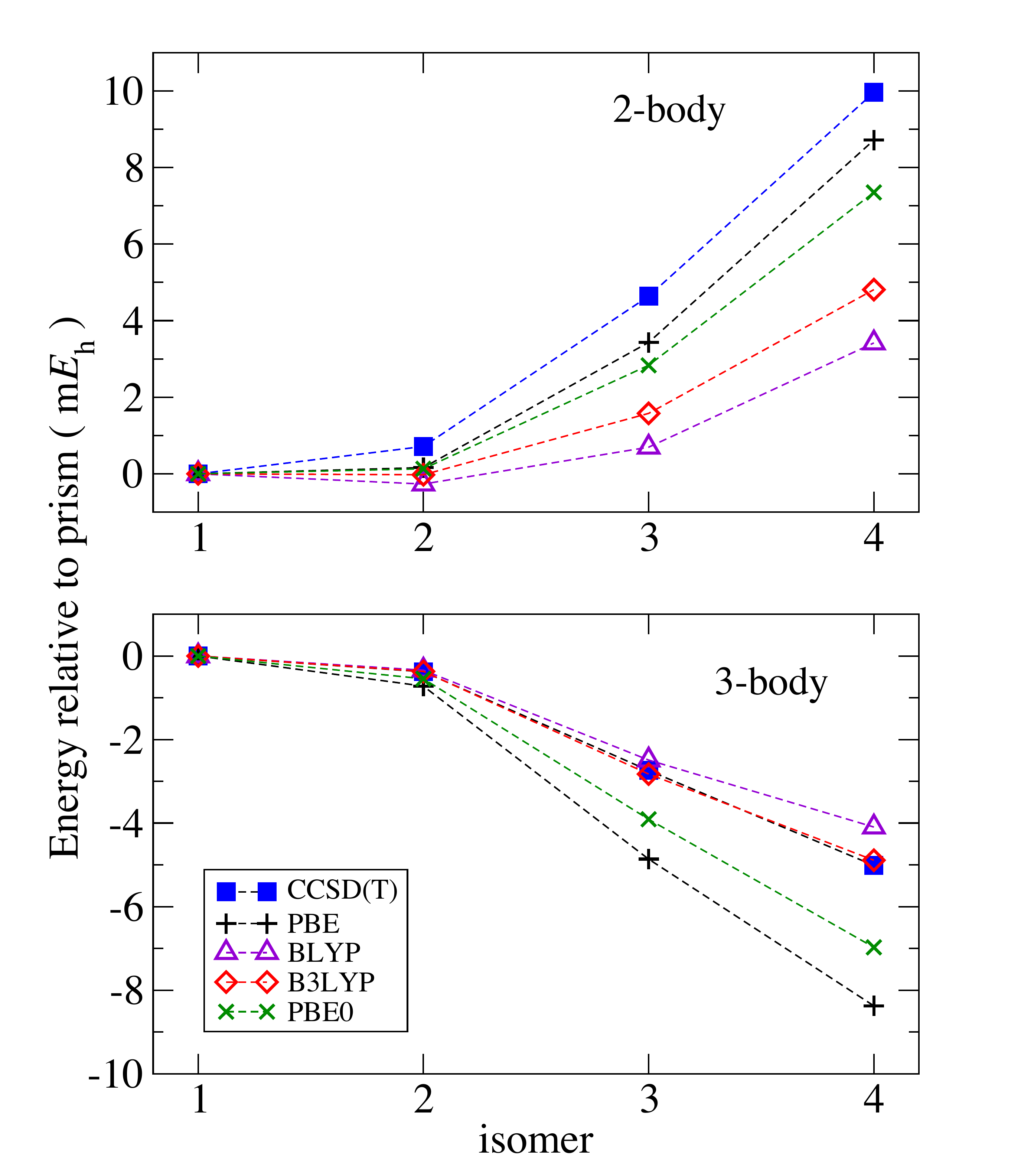}}
\caption{Comparison of DFT values of 2-body energies (upper panel) and
3-body energies (lower panel) of four isomers
of the H$_2$O hexamer with benchmark values from CCSD(T). Numbering
of isomers is prism: 1, cage: 2, book: 3, ring: 4.
Units: m$E_{\rm h}$}
\label{fig:he_det2b_3b_bench_dft_1-4}
\end{figure}

To understand the origin of the erroneous DFT predictions, we compare in 
Fig.~\ref{fig:he_det2b_3b_bench_dft_1-4}
the 2-body and 3-body energies from benchmark CCSD(T)
and from DFT. The 2-body benchmark energies were obtained 
using MP2-F12/AVQZ
and $\Delta$CCSD(T)-F12/AVTZ, while for the 3-body benchmarks we used AVTZ
for all parts. The DFT 2- and 3-body energies were computed using AVQZ
and AVTZ respectively. It is clear from the 2-body
results that BLYP and B3LYP predict
much too weak a lowering of 2-body energy as we go from the ring to
the cage and prism, while PBE is rather accurate and PBE0 is less accurate
than PBE but better than BLYP and B3LYP.
(The difference of 2-body energies of ring and prism
is $10.0$~m$E_{\rm h}$
according to CCSD(T), but is only $3.4$ and $4.7$~m$E_{\rm h}$
with BLYP and B3LYP respectively, so that BLYP is in error by
a factor of 3 and B3LYP by a factor of 2.)
This is what we should expect from the 2-body
energies presented in Sec.~\ref{sec:dimer_thermal}, 
since BLYP and B3LYP are systematically
too repulsive over a rather wide range of distances, while the errors
of PBE and PBE0 are much smaller.
By contrast, in the 3-body energy the situation is reversed, with
B3LYP now being almost perfect and BLYP only a little worse,
while PBE and PBE0 are both rather poor. (The ring-prism difference
of 3-body energy is $5.01$~m$E_{\rm h}$ according to CCSD(T), 
but  PBE and PBE0 give $8.38$ and $6.97$~m$E_{\rm h}$ respectively.)
We expect from this that if we correct
the DFT approximations for their 1- and 2-body errors, thus obtaining
what we referred to earlier as DFT-\deltatwo, then BLYP-\deltatwo\ and
B3LYP-\deltatwo\ 
should agree rather well with CCSD(T) and DMC, while PBE-\deltatwo\
and PBE0-\deltatwo\ 
should be less good. This expectation is confirmed by our DFT-\deltatwo
energies reported in Table~\ref{tab:hex_isomer}. If we now correct
also for 3-body errors, we would expect this to give little further
improvement for BLYP and B3LYP, but substantial improvements
for PBE and PBE0, and the DFT-\deltathree\ results of Table~\ref{tab:hex_isomer}
confirm this.

These comparisons are very useful for our assessment of the
accuracy of DMC, because they indicate that DMC must be accurate
not only for the 2-body energy, as we already know from Sec.~\ref{sec:dimer},
but also for the 3-body energy. We can draw this conclusion because PBE
and DMC have very similar, and very small errors for 2-body energy,
but DMC is very much better than PBE for the hexamer isomers, and we
have traced the main cause of this to the 3-body energy.

\section{Discussion and conclusions}
\label{sec:discussion}

The main aims of the present work have been to assess the accuracy
of diffusion Monte Carlo (DMC) 
for water clusters by comparing with quantum chemistry benchmarks,
and to investigate how well it overcomes the deficiencies of common
DFT approximations. We noted in the Introduction that any method
that is intended to give an accurate description of cluster and bulk water
and ice systems across a wide range of conditions must be accurate for
all the key components of the energy, including the distortion
energy of the H$_2$O monomer, the 2-body interactions that
determine the energetics of the water dimer, and the many-body
contributions arising from polarisability and other mechanisms
that are known to be crucial for larger clusters and for the bulk
liquid and solid phases. Our comparisons with well converged CCSD(T)
energies for both statistical samples of configurations and in some cases
for sets of stable isomers show that DMC gives all the main
components of the energy rather accurately, while
the standard DFT approximations that we have studied encounter problems with
one or more of these components. 
We have emphasised the importance of studying random thermal samples,
since these allow us to characterise the accuracy
of QMC and DFT approximations across an entire domain of configurations,
rather than at a small number of special configurations. At the same time,
we have noted that the mean and rms errors of any given approximation
will depend on the choice of thermal sample.

The importance of achieving an accurate description of monomer
energetics was emphasised in a recent paper 
of Santra {\em et al.}~\cite{santra09},
who showed that some commonly used DFT functionals 
gives a rather poor description of the distortion
energy, making bond-stretching too easy. We have confirmed this on a thermal
sample of 100 distorted monomer configurations drawn from a classical m.d.
simulation of liquid water performed with the flexible {\sc amoeba} 
interaction model. We found poor accuracy with PBE and BLYP, better accuracy
with B3LYP and excellent accuracy with PBE0, in 
agreement with Ref.~\cite{santra09}.
The accuracy of DMC turns out to be even better than PBE0. It has
been suggested that the excessive deformability of the H$_2$O monomer
with PBE and BLYP may be a significant factor in their rather poor
predictions for bulk water. The very high accuracy of DMC for the monomer
means that it does not suffer from such problems.

Our calculations on the 10 stationary points of the H$_2$O dimer
provide important evidence that the accuracy of 
DMC for the 2-body energy of water systems
is also very good. All the energies come in the correct order, though we
recall that this is also achieved by most DFT 
approximations. Much more significant
is the very close agreement with highly converged CCSD(T) benchmarks
for the energies relative to the global minimum, which are almost all
reproduced by DMC to within $0.1$~m$E_{\rm h}$ ($2.7$~meV). By contrast,
DFT errors for the relative energies are 
typically $0.5$~--~$1.0$~m$E_{\rm h}$, and no
DFT approximation that we are aware of comes near the accuracy of DMC.
We note that these comparisons give a direct test of the 2-body
energy, since the monomer distortion energies at the 10 stationary
points are extremely small.

More relevant to bulk-phase water are our comparisons between DMC, DFT
and CCSD(T) benchmarks for a large random thermal sample 
of dimer configurations drawn
from the {\sc amoeba} m.d. simulation of bulk water. This sample
is large enough for us to examine errors as a function of O-O
distance, and we have seen that DMC reproduces the benchmarks accurately
and consistently throughout the range $2.5$~-- $7.0$~\AA\ that we have
examined. The DMC errors barely exceed
the statistical errors of the Monte Carlo sample of the DMC calculations
themselves, the mean and rms deviation of the DMC energy from the
CCSD(T) benchmarks being 0.018 and 0.102~m$E_{\rm h}$ (0.5 and 2.8~meV).
These errors are of about the size that might be expected
from previous DMC work on the H$_2$O dimer. On the other hand, the
errors of the total dimer energy with all the DFT approximations
examined here are very much greater (see 
Table~\ref{tab:dimer_errors}). However, in the case
of PBE and BLYP, the errors in the monomer distortion energy
contribute significantly. In practical calculations using these 
approximations on clusters or bulk systems, it would be perfectly
straightforward to correct for these 1-body errors, simply by adding
the difference between the essentially exact Partridge-Schwenke
and the DFT distortion energies. If we do this for our dimer samples,
then we obtain corrected DFT approximations which we refer to as DFT-\deltaone,
whose errors are solely in the 2-body part. We have
seen that the 2-body energy of BLYP is much too repulsive and B3LYP
suffers from the same problem, as would be expected from the
substantial underbinding of the dimer with BLYP and B3LYP.
However, for PBE, the 2-body energy turns out to be very good, its
quality being comparable with that of DMC, so that if we simulated
the dimer with PBE corrected for 1-body errors, rather accurate
results would be obtained; the approximation PBE0-\deltaone\ is also quite
respectable. 

For the larger clusters from the trimer to the hexamer, we have
followed a similar procedure, drawing sets of configurations
at random from the {\sc amoeba} m.d. simulation and comparing
DMC, DFT and benchmark CCSD(T) energies for these samples,
taking care as usual that the total energies with DFT and CCSD(T)
are basis-set converged to $\sim 0.1$~m$E_{\rm h}$ or better. For these
clusters, the thermal samples are smaller than for the dimer, consisting
of 50 configurations for the trimer and 25 each for the tetramer,
pentamer and hexamer. Since the errors in either or both of the
1-body and 2-body components with the DFT approximations are
considerably larger than those of DMC, we expect that DMC will
substantially outperform them for these large clusters, and this is
indeed what we find. However, this is not the whole story, because
it is possible that some of the
problems encountered by DFT approximations in treating bulk liquid
water and ice may be associated with many-body (i.e. beyond-2-body)
components of the energy, perhaps because their description of
the relevant polarisabilities is inadequate. We have therefore
tried to test whether DMC also outperforms DFT for these many-body
contributions.

One way we have used to test the quality of the DMC beyond-2-body
energy is based on making a many-body decomposition of the DFT
total energy for our thermal samples of the trimer and higher clusters,
and then to replace the DFT 1- and 2-body energies by the benchmark
energies (i.e. Partridge-Schwenke for 1-body and near-CBS CCSD(T)
for 2-body). Any remaining errors in the resulting corrected versions
of DFT, which we refer to as DFT-\deltatwo, are then due entirely to errors
in the beyond-2-body energy. If we then
compare with DMC, we are putting DMC to an extreme and certainly unfair test,
since it has to compete unaided against massively assisted DFT.
Remarkably, DMC survives even this rather well, having errors in its total
energy that are still smaller than or comparable with
the errors in the beyond-2-body energy for the DFT approximations.

We have noted that the relative energies of the well-known
isomers of the H$_2$O hexamer also provide an excellent way of
testing the beyond-2-body energy of DMC. The point here is that
all the DFT approximations we examined give completely erroneous
energy orderings of these isomers. It has been shown 
in earlier work~\cite{santra08}
that DMC gives the energy differences between these isomers in 
excellent agreement with CCSD(T), and we confirmed this here by comparing
with the improved CCSD(T) energies now available. As also
pointed out earlier, a many-body decomposition of the DFT and CCSD(T)
energies allows one to determine where the DFT errors come from. We have
presented our own many-body analysis showing that for some of the
DFTs (e.g. BLYP) the errors lie mainly in the 2-body part, whereas
in others (e.g. PBE) the 2-body component is accurate, but there are
substantial errors in the beyond-2-body components. Since we know
that DMC gives an accurate account of the 2-body component,
these comparisons confirm its accuracy also for the beyond-2-body
components.

It is intriguing that the isomers of the hexamer reveal the superiority
of DMC over some of the DFTs for beyond-2-body energy in a much clearer
way than the thermal sample of hexamer configurations. The implication
of this is that thermal samples of cluster configurations drawn from
a realistic model of the liquid do not necessarily suffice for a full
assessment of the errors of approximate methods. It is instructive to
note that if an isolated hexamer in free space were simulated in thermal
equilibrium using one of our DFT approximations, a completely erroneous
distribution consisting mainly of ring-like structures would be observed,
whereas if the simulation were performed with DMC (assuming this to be
feasible), a much more realistic distribution consisting mainly
of compact structures would be observed. However, the thermal sample
of configurations generated by DMC would not suffice to assess fully
the errors of DFT, because these errors only become apparent for thermal
samples that include both open and compact structures. Similarly in 
assessing DFT errors using thermal samples relevant to the liquid, 
it seems desirable to use a wider range of configurations than those
that occur commonly in the real liquid. This is an important matter
for future study.

The present comparative study on water clusters, taken together with
the recently demonstrated high accuracy of DMC for the
energetics of several ice crystal structures~\cite{santra11}, 
indicates that DMC has the
accuracy needed to serve as a benchmarking tool for water systems
across a wide range of conditions. Its advantages over correlated
quantum chemistry techniques are that its scaling with system
size is much more favourable, convergence to the basis-set limit
is easily achieved, it is straightforward to apply to periodic
systems, and its parallel scaling on large supercomputers is essentially perfect. 
In the immediate future, we plan to use DMC to obtain 
benchmark energies for thermal samples of much larger water clusters
than those studied here. These benchmarks will then be used to test
DFT approximations. We expect to find ways of separating both the
benchmark and the DFT total energies into their 1-, 2- and beyond-2-body
contributions, as we have done here, so that we can analyse the origin
of DFT errors. It should also be possible to do the same for
bulk water itself. As we noted in the Introduction, DMC calculations
on thermal samples of periodic liquid water configurations were already
demonstrated some years ago~\cite{grossman05}, 
so that the technical feasibility of
what we suggest is not in doubt. The possibility of tuning
DFT approximations to reproduce DMC and quantum chemistry benchmarks for
a range of water systems is also an interesting possibility for the future.
It is worth adding that tests of DMC against CCSD(T)
for larger clusters should be
possible in the future, because the availability of quantum chemistry
codes that can be run on very large parallel computers is already
making it possible to perform coupled-cluster calculations
on much larger molecular systems than before~\cite{yoo10}.

In conclusion, we have shown that the accuracy of DMC for thermal and
other con\-figuration samples of H$_2$O clusters from the
monomer to the hexamer is
excellent, the errors of DMC being typically 
$0.1 - 0.2$~m$E_{\rm h}$ ($2.5 - 5.0$~meV) per
molecule. This error is much smaller than the typical DFT errors,
and this, together with other evidence for the
accuracy of DMC, supports its reliability as a source of benchmarks
for testing and calibrating DFT. It is desirable to extend the tests
of DMC accuracy to larger clusters, and we have indicated how this
might be possible. We have also suggested how DMC calculations 
can be used to improve the
understanding of both liquid and solid bulk water.


\section*{Acknowledgments}

The work of MDT is supported by EPSRC grant EP/I030131/1, and that of FRM by EPSRC grant EP/F000219/1.
DMC calculations
were performed on the Oak Ridge Leadership Computing Facility, located
in the National Center for Computational Sciences at Oak Ridge
National Laboratory, which is supported by the Office of Science of
the Department of Energy under contract DE-AC05-00OR22725 (USA).



\clearpage

\section*{Tables}

\begin{table}[!htb]
\begin{center}
\begin{tabular}{l|cc}
\hline
   & mean & rms \\
\hline
DMC   & $0.01$ & $0.04$ \\
PBE   & $-0.37$ & $0.54$ \\
BLYP  & $-0.45$ & $0.63$ \\
B3LYP & $-0.14$ & $0.20$ \\
PBE0  & $-0.01$ & $0.08$ \\
\hline
\end{tabular}
\end{center}
\caption{Mean values and rms fluctuations of DMC and DFT errors of monomer
energy for a thermal sample of 100 configurations (see text).
The Partridge-Schwenke (PS) energy function is used as the ``exact'' energy,
and the energy zero for DMC and DFT approximations is taken to
be the energy in the PS equilibrium geometry. Units: m$E_{\rm h}$.}
\label{tab:monomer_energy}
\end{table}


\begin{table}[!htb]
\begin{center}
\begin{tabular}{lcccccc}
\hline
  s.p. &  CCSD(T) &  DMC  & PBE & BLYP & B3LYP & PBE0 \\
\hline
   1  &  0.000  & 0.000  & 0.000  & 0.000  & 0.000  & 0.000  \\
   2  &  0.783  & 0.866  & 0.904  & 0.838  & 0.834  & 0.867  \\
   3  &  0.910  & 1.001  & 1.288  & 1.194  & 1.040  & 1.085  \\
   4  &  1.115  & 1.250  & 1.707  & 1.814  & 1.690  & 1.627  \\
   5  &  1.519  & 1.445  & 2.374  & 2.406  & 2.208  & 2.187  \\
   6  &  1.603  & 1.505  & 2.707  & 2.676  & 2.382  & 2.401  \\
   7  &  2.893  & 2.863  & 3.544  & 3.534  & 3.447  & 3.395  \\
   8  &  5.663  & 5.739  & 5.869  & 5.698  & 5.881  & 5.883  \\
   9  &  2.840  & 2.910  & 3.502  & 3.485  & 3.371  & 3.304  \\
   10 &  4.307  & 4.471  & 4.838  & 4.591  & 4.601  & 4.654  \\
\hline
\end{tabular}
\end{center}
\caption{Comparison of DMC energies and DFT energies given by
the PBE, BLYP, B3LYP and PBE0 functionals with CCSD(T) benchmarks
for the 10 stationary points of the H$_2$O dimer. For each set of
energies, the zero of energy has been taken so that the energy
of the global minimum geometry is equal to zero. Numbering
of stationary points follows that of previous authors
(see e.g. Refs.~\cite{tschumper02}). Energy units: m$E_{\rm h}$.}
\label{tab:sp_bench_DMC_DFT}
\end{table}


\begin{table}[!htb]
\begin{center}
\begin{tabular}{lcc}
\hline
  method & mean error  & rms error  \\
\hline
DMC  & 0.018  & 0.102  \\
PBE  & -0.681  & 0.783  \\
BLYP   & -0.155  & 1.102  \\
B3LYP   & 0.180  & 0.463  \\
PBE0    & 0.093  & 0.177  \\
PBE-\deltaone   & 0.056  & 0.144  \\
BLYP-\deltaone   & 0.736  & 0.679  \\
B3LYP-\deltaone   & 0.469  & 0.371  \\
PBE0-\deltaone   & 0.111  & 0.133  \\
\hline
\end{tabular}
\end{center}
\caption{DMC and DFT mean and rms fluctuation of errors of the total
energy for thermal sample of 198 dimer configurations.
In the case of DFT, the approximations denoted by PBE-\deltaone\ etc
represent the total energy after correction for errors of the 1-body
energy (see text). Units: m$E_{\rm h}$.}
\label{tab:dimer_errors}
\end{table}


\begin{table}[!htb]
\begin{center}
\begin{tabular}{lrrr}
\hline
DFT & \multicolumn1c{DFT}  & \multicolumn1c{DFT-\deltaone} & \multicolumn1c{DFT-\deltatwo} \\
\hline
\multicolumn{4}{c}{trimer} \\
PBE&$-$0.034	(0.376)&	0.152	(0.137)&	0.062	(0.049) \\
BLYP&0.517	(0.413)&	1.101	(0.256)&	$-$0.023	(0.039) \\
B3LYP&0.542	(0.175)&	0.709	(0.147)&	$-$0.011	(0.028) \\
PBE0&0.215	(0.132)&	0.212	(0.125)&	0.031	(0.026) \\		
\multicolumn{4}{c}{DMC: $0.100$ ($0.043$)} \\
\hline
\multicolumn{4}{c}{tetramer} \\
PBE&$-$0.163	(0.231)&	0.230	(0.143)&	0.110	(0.070) \\
BLYP&0.861	(0.392)&	1.331	(0.335)&	$-$0.041	(0.036) \\
B3LYP&0.709	(0.219)&	0.859	(0.206)&	$-$0.019	(0.025) \\
PBE0&0.272	(0.126)&	0.282	(0.119)&	0.056	(0.038) \\
\multicolumn{4}{c}{DMC: $0.105$ ($0.044$)} \\
\hline
\multicolumn{4}{c}{pentamer} \\
PBE&$-$0.153	(0.282)&	0.276	(0.125)&	0.156	(0.077) \\
BLYP&1.191	(0.335)&	1.703	(0.215)&	$-$0.101	(0.050) \\
B3LYP&0.902	(0.168)&	1.060	(0.152)&	$-$0.056	(0.035) \\
PBE0&0.297	(0.121)&	0.304	(0.131)&	0.075	(0.042) \\
\multicolumn{4}{c}{DMC: $0.154$ ($0.046$)} \\
\hline
\multicolumn{4}{c}{hexamer} \\
PBE&$-$0.068	(0.266)&	0.318	(0.134)&	0.178	(0.077) \\
BLYP&1.298	(0.376)&	1.754	(0.304)&	$-$0.107	(0.053) \\
B3LYP&0.989	(0.180)&	1.123	(0.175)&	$-$0.058	(0.037) \\
PBE0&0.367	(0.114)&	0.367	(0.112)&	0.088       (0.041) \\
\multicolumn{4}{c}{DMC: $0.183$ ($0.044$)} \\
\hline
\end{tabular}
\end{center}
\caption{Deviations of energies given by corrected DFT
  approximations away from CCSD(T) benchmark energies for
  thermal samples of (H$_2$O)$_n ~ 3 \le n \le 6$ configurations. Notations DFT, DFT-\deltaone\ and
  DFT-\deltatwo\ indicate DFT approximations uncorrected, corrected
  for 1-body errors, and corrected for both 1- and 2-body errors. Each entry gives the mean
  deviation, with rms fluctuation of the
  deviation in parentheses. Mean and rms deviations of
  DMC energies away from CCSD(T) benchmarks are shown
  for comparison. Energy units: m$E_{\rm h}$ per water monomer.}
\label{tab:compare_DFT-n}
\end{table}


\begin{table}[!htb]
\begin{center}
\begin{tabular}{l|cccc}
\hline
 method & prism & cage & book & ring \\
\hline
 MP2         & $0.00$ & $0.08$ & $0.51$ & $1.84$ \\
 MP2$^\dagger$ & $0.00$ & $0.10$ & $0.53$ & $1.93$ \\
 CCSD(T)     & $0.00$ & $0.39$ & $1.12$ & $2.70$ \\
 CCSD(T)$^\dagger$ & $0.00$ & $0.40$ & $1.15$ & $2.87$ \\
 DMC         & $0.00$ & $0.53$ & $0.90$ & $2.45$ \\
\hline
 PBE & $0.00$ & $-0.55$ ($-0.95$) & $-1.99$ ($-3.14$) & $-1.45$ ($-4.32$) \\
 PBE-\deltatwo & $0.00$ & $0.00$ ($-0.40$) & $-0.78$ ($-1.93$) & $-0.20$ ($-3.07$) \\
 PBE-\deltathree & $0.00$ & $0.35$ ($-0.05$) & $1.34$ ($0.19$) & $3.16$ ($0.29$) \\
       &        &                  &                 &                 \\
 BLYP & $0.00$ & $-0.94$ ($-1.34$) & $-3.47$ ($-4.62$) & $-3.90$ ($-6.77$) \\
 BLYP-\deltatwo & $0.00$ & $0.04$ ($-0.36$) & $0.47$ ($-0.68$) & $2.64$ ($-0.23$) \\
 BLYP-\deltathree & $0.00$ & $0.01$ ($-0.39$) & $0.22$ ($-0.93$) & $1.72$ ($-1.15$) \\
       &        &                  &                 &                    \\
 B3LYP & $0.00$ & $-0.69$ ($-1.09$) & $-2.65$ ($-3.80$) & $-2.98$ ($-5.85$) \\
 B3LYP-\deltatwo & $0.00$ & $0.05$ ($-0.35$) & $0.41$ ($-0.74$) & $2.17$ ($-0.70$) \\
 B3LYP-\deltathree & $0.00$  & $0.04$ ($-0.36$) & $0.49$ ($-0.66$) & $2.04$ ($-0.83$) \\
       &        &                  &                 &                      \\
 PBE0 & $0.00$ & $-0.45$ ($-0.85$) & $-1.76$ ($-2.91$)& $-1.65$ ($-4.52$) \\
 PBE0-\deltatwo & $0.00$ & $0.13$ ($-0.27$) & $0.04$ ($-1.11$) & $0.96$ ($-1.91$) \\
 PBE0-\deltathree & $0.00$ & $0.30$ ($-0.10$) & $1.21$ ($0.06$) & $2.92$ ($0.05$) \\
\hline
\end{tabular}
\end{center}
\caption{Total energies of selected isomers of the water hexamer
relative to that of the prism, calculated by different methods.
In all cases, the geometry of the isomer is the relaxed geometry
given by MP2 calculations with the AVTZ basis, as given in the Supplementary
Information of Ref.~\cite{santra08}. All energies 
were calculated in the present work, except
for the MP2 and CCSD(T) energies marked with $\dagger$ from Ref.~\cite{bates09}
and the DMC energies from Ref.~\cite{santra08}. 
Entries DFT-$n$ with $n = 2$ and $3$ are DFT energies 
corrected for 1- and 2-body errors, and corrected for 1-, 2- and 3-body
errors respectively. Values in parentheses
represent errors compared with the CCSD(T) energies from Ref.~\cite{bates09}. 
Energy units: m$E_{\rm h}$.}
\label{tab:hex_isomer}
\end{table}


\clearpage

\section*{Figures}

\end{document}